\documentclass[prd,preprint]{revtex4}

\newcommand{\h}{\mathcal{H}}
\newcommand\R{\zeta}

\newcommand{\geffs}{G_{\rm eff}^\Psi} 
\newcommand{\geffsh}{G_{\rm eff}^{\Psi+\Phi}}

\def\MM{M_{\rm P}}

\newcommand\Lr{\rm L}

\usepackage{xcolor}

\usepackage[utf8]{inputenc}

\usepackage{latexsym}
\usepackage{epsfig}
\usepackage{graphicx,subfigure}
\usepackage[normalem]{ulem}
\usepackage{color}
\usepackage{xcolor}
\usepackage{multirow}
\usepackage{hyperref}
\usepackage{numprint}
\usepackage{eucal}
\usepackage{amsmath}
\usepackage{amssymb}

	\usepackage{amsmath}
	\usepackage{amsfonts}
  	\usepackage{amssymb}
	\usepackage{makeidx}
	\usepackage{amsfonts}
	\usepackage[ansinew]{}
	\usepackage[usenames,dvipsnames]{pstricks}
	\usepackage{epsfig}

    \usepackage{graphicx}

	\usepackage{float}

\newcommand{\be}{\begin{equation}}
\newcommand{\ee}{\end{equation}}
\newcommand{\bea}{\begin{eqnarray}}
\newcommand{\eea}{\end{eqnarray}}
\newcommand{\ba}{\begin{align}}
\newcommand{\ea}{\end{align}}

\makeindex

\begin{document}

\title{Multimessenger consistency relations bridging gravitational wave and large scale structure observations}

\author{Antonio Enea Romano}

\affiliation{ICRANet, Piazza della Repubblica 10, I--65122 Pescara, Italy}

\begin{abstract}
    We show that for Horndeski theories it is possible to derive mathematically compact consistency relations (CR) between physically observable quantities, valid for different classes of theories defined by the behavior of the brading  function $\alpha_B$, independent of all other property functions. 
    The CRs establish a parametrization independent direct relation between the effective gravitational constant, the slip parameter, the gravitational and electromagnetic waves (EMW) luminosity distances, the speed of gravitational waves (GW) and the sound speed. The no-brading CR is also satisfied by general relativity (GR), and allows to estimate the gravitational coupling from GWs observations, independently from large scale structure (LSS) observations. A general, less mathematically compact, consistency condition is also derived, valid for any form of the function $\alpha_B$ and the other property functions.
    
    We apply the CRs to map the large scale structure  observational constraints on the effective gravitational constant and the slip parameter to GW-EMW distance ratio constraints, showing that LSS and GWs give independent constraints consistent with no-brading. Beside allowing to perform  parametrization and model independent tests of the consistency between different constraints on modified gravity,  the CRs allow to probe the value of the effective gravitational constant with multimessenger observations, independently from LSS observations. 
\end{abstract}


\maketitle

\section{Introduction}
The detection of gravitational waves (GWs) \cite{LIGOScientific:2016aoc} allows to test general relativity and its possible modifications. 
The effects of modified gravity do not only affect GWs but also other physical phenomena, such as for example large scale structure formation, and is for this reason important to investigate the consistency between these different effects.
Modified gravity theories are often studied assuming some phenomenological ansatzes, however this can sometime lead to a misestimation of the  
observables \cite{Linder:2016wqw}, and depend on the form of the adopted parametrization.
For this reason it is important to develop parametrization independent tests of modified gravity, relating directly physical observables. This would allow to assess the compatibility with observational data of large classes of theories, without making any assumption about the functions defining  them.

In this paper we show that for different large classes of modified gravity theories the EFT of dark energy \cite{Gubitosi:2012hu} allows to derive  consistency relations between the effective gravitational constant, the slip parameter, the gravitational and electromagnetic luminosity distance, the speed of GWs and the sound speed. These CRs  generalize  the results obtained in some luminal modified gravity theories. We apply the consistency relations to obtain the GW-EMW distance ratio constraints implied by large scale structure observations, showing that they are consistent with no-brading.

\section{Effective theory}
The quadratic effective field theory action (EFT) of perturbations for a single scalar  dark energy field was derived in \cite{Gleyzes:2013ooa} 

\be
\begin{split}
\label{total_action}
S = \int \! d^4x \sqrt{-g} \Bigg[ \frac{\MM^2}{2} f(t) R - \Lambda(t) - c(t) g^{00}  + \, \frac{M_2^4(t)}{2} (\delta g^{00})^2\, 
-\, \frac{m_3^3(t)}{2} \, \delta K \delta g^{00} -\, \\ m_4^2(t)\left(\delta K^2 - \delta K^\mu_{ \ \nu} \, \delta K^\nu_{ \ \mu} \right) + \frac{\tilde m_4^2(t)}{2} \, \R \, \delta g^{00}  \Bigg] \,,
\end{split}
\ee
where $\zeta$ is the curvature perturbation, $K_{\mu \nu}$ is the extrinsic curvature tensor, $\delta g^{00} \equiv g^{00} +1$, $\delta K_{\mu \nu} \equiv K_{\mu \nu} - H h_{\mu \nu}$, and $K \equiv K^{\mu}_{\ \mu}$ and $M_P$ is the Planck mass.
The above action for tensor modes gives \cite{Gubitosi:2012hu,Gleyzes:2013ooa}
\be
S_{\gamma}^{(2)} =\int d^4 x \, a^3 \frac{\MM^2 f }{v^2_{\rm GW}} \left[   \dot{\gamma}_{ij}^2 -\frac{v^2_{\rm GW}}{a^2}(\partial_k \gamma_{ij})^2 \right]\,, \label{lgamma}
\ee
where the GWs speed is related to the EFT action coefficients by
\be
v_{\rm GW}^2 =  \left(1+\frac{2m_4^2}{\MM^2 f}\right)^{-1}\;.\label{vEFT}
\ee

\section{EFT reconstruction of Horndeski theories}
The EFT of dark energy can be applied to a very large class of theories \cite{Bloomfield:2012ff},  it allows to compare different theories in terms of the EFT coefficients, and to conveniently perform model independent analysis of observational data. The relation between the EFT coefficients and fundamental covariant theories, such as Horndeski theories, is not trivial \cite{Kennedy:2017sof}, and once established it allows to interpret constraints on  EFT coefficients in terms of viable classes of covariant  theories. 
The advantage of this approach consists in deriving general model independent results based on the EFT, instead of having to make theory specific derivations, and then to apply them to specific families of theories. The correspondence between EFT and $\alpha$ functions for Horndeski theories is shown in Table \ref{EFTalpha}.
\begin{table*}[t]
\centering
\begin{tabular}{|c|c|c| }
\hline
  EFT functions  &$\alpha$-parametrization \\
\hline
& \\ [-1.5ex]
 $f(t)$ &$\frac{M_{*}^{2}}{\MM^{2}}c_{T}^{2}$ \\[3ex]
 $\frac{2c(t)}{\MM^{2}}$  &$-\frac{\rho_{m}}{\MM^{2}}-\frac{M_{*}^{2}}{\MM^{2}}\beta(t)    $ \\ [3ex]
 $\frac{\Lambda(t)-c(t)}{\MM^{2}}$& $\frac{M_{*}^{2}}{\MM^{2}}\left[3H^{2}c_{T}^{2}(1+\alpha_{M})+\beta(t)+3H\dot{\alpha}_{T}\right]$   \\ [3ex]
$M_{2}^{4}(t)$ & $\frac{1}{4}\rho_{m}+\frac{M_{*}^{2}}{4}\left[H^{2}\alpha_{K}+\beta(t) \right]$ \\ [3ex]
$m_{3}^{3}(t)$ &$M_{*}^{2}\left[H\alpha_{M}c_{T}^{2}+\dot{\alpha}_{T}-2H\alpha_{B}   \right]  $ \\ [3ex]
$m_{4}^{2}(t)$ & $-\frac{1}{2}M_{*}^{2}\alpha_{T}$ \\ [1.5ex]
\hline
\end{tabular}
\caption {Relationship \cite{Kennedy:2017sof} between the EFT functions adopted in \cite{Gubitosi:2012hu} and the $\alpha$ property functions defined in \cite{Bellini:2014fua}.
 The dots denote derivatives with respect to time,  and the  function $\beta$ is defined by $\beta(t) \equiv c_{T}^{2}\left[2\dot{H}+H\dot{\alpha}_{M}+\alpha_{M}\left(\dot{H}-H^{2}+H^{2}\alpha_{M}\right)\right]+ H\dot{\alpha}_{T} (2\alpha_{M}-1)+\ddot{\alpha}_{T}$. } 
 \label{EFTalpha}
\end{table*}

For GWs observations the general EFT prediction for the GW-EMW distance ratio was derived in \cite{Romano:2023xal,Romano:2023ozy}. These model independent results confirm and generalize to non-luminal theories the results obtained previously in some specific classes of theories.
For large scale structure observations the general EFT predictions for the effective gravitational coupling and the slip parameter were computed in \cite{Linder:2015rcz}, and were then applied to Horndeski theories, expressing them in terms of the $\alpha$ property functions \cite{Bellini:2014fua}.

In this paper we will combine together these different theoretical results to obtain consistency relations between  GWs and LSS observables, eliminating the property functions dependency as much as possible.
This approach allows to obtain theoretical relations between physical observable, limiting the reliance on specific parameterizations used to analyze observational data, and to compare results obtained using different parameterizations. Note that this can be important in analyzing observational data, because commonly used parameterizations for the $\alpha$ functions can lead to the misestimation of observables \cite{Linder:2016wqw}. 
Beside this, it is useful do derive compact model independent analytical consistency relations which help to gain physical intuition about the general interconnection between GWs and LSS observables.

\section{Effects of modified gravity on gravitational waves}
Note that $v_{\rm GW}$ depends on the ratio of two coefficients of the EFT action, $m_4$ and $f$, so that observational constraints on $v_{\rm GW}$ are  mapped into constraints of this ratio, not of the individual coefficients of the action. 
We can conveniently rewrite the effective action in eq.(\ref{lgamma}) as
\be
S_{\gamma}^{(2)} =\int d^4 x \,  \frac{a^2 \Omega^2}{v^2_{\rm GW}} \left[   {\gamma'}_{ij}^2 -v^2_{\rm GW}(\partial_k \gamma_{ij})^2 \right]=\int d^4 x \,\alpha^2\Big[ \gamma_{ij}'^2-v_{\rm GW}^2 (\partial_k \gamma_{ij})^2\Big]~, \label{lgammaeta2}
\ee
which gives the equation of motion \cite{Romano:2023xal}
\be
\gamma_{ij}''+2  \frac{\alpha'}{\alpha} \gamma_{ij}'-v^2_{\rm GW} \nabla^2 \gamma_{ij}=
\gamma_{ij}''+2  \h\Big(1-\frac{v_{\rm GW}'}{\h v_{\rm GW} }+\frac{\Omega'}{\h\Omega}\Big) \gamma_{ij}'-v^2_{\rm GW} \nabla^2 \gamma_{ij}=0  \,,\label{heft}
\ee
where we have introduced 
\be
\alpha=\frac{a\, \Omega}{v_{\rm GW}}\,,
\ee
and prime denotes derivative with respect to conformal time.
The WKB approximation of the solution of the propagation equation gives
\bea
\gamma_{ij}
&\propto& \frac{1}{r\,\alpha  \sqrt{v_{\rm GW}}}=\frac{\sqrt{v_{\rm GW}}}{r\,\Omega  } \,
\eea
which implies that the GW amplitude scales as
\be
\frac{\gamma_{ij,o}}{\gamma_{ij,e}}\propto \frac{\alpha_e}{\alpha_o}\sqrt{\frac{v_{\rm GW,e}}{v_{\rm GW,o}}}=\frac{a_e}{a_o}\frac{\Omega_e}{\Omega_o} \sqrt{\frac{v_{\rm GW,o}}{v_{\rm GW,e}}}\,, \label{hMGT}
\ee
where the subscripts $e$ and $o$ denote respectively  the emitted and observed quantities. 

Taking into account the  relation between the emitted and observed GWs amplitude the GW-EMW distance ratio is then given by \cite{Romano:2023xal}  
\be
r_d(z)=\frac{d^{\rm GW}_{\Lr}(z)}{d^{\rm EM}_{\Lr}(z)}=\frac{M_{eff}(0)}{M_{eff}(z)} \sqrt{\frac{v_{\rm GW}(z)}{v_{\rm GW}(0)}}=\sqrt{ \frac{f(0)v_{\rm GW}(z)}{f(z)v_{\rm GW}(0)}}=\frac{M_*(0)}{M_*(z)} \sqrt{\frac{v_{\rm GW}(0)}{v_{\rm GW}(z)}}   \,,\label{rd}
\ee
where we have  defined $M_*=M_P \Omega/v_{\rm GW}$.
Note that for GWs the observationally relevant parameter is $f$, not its time derivative $\alpha_M$. 

\subsection{Observational constraints}
The most precise constraints on the GW-EMW ratio $r_d(z)$ are imposed by bright sirens, GWs events with an electromagnetic counterpart. 
The only bright siren event observed so far,  GW170817 \cite{LIGOScientific:2017zic}, gives the constraint $r_d(z\approx0.001)=0.89^{+0.12}_{-0.2}$, which is consistent with the GR prediction $r_d=1$.

For dark sirens, GW events without an electromagnetic counterpart, a joint estimation of cosmological and modified gravity parameters can be performed \cite{Chen:2023wpj}. Using the parametrization \cite{Belgacem:2017ihm}
\be
r_d(z)=\Xi_0+\frac{1-\Xi_0}{(1+z)^n} \label{Xi0}\,,
\ee
the best fit parameters inferred analyzing GWs emitted by black holes and neutron stars binary systems  \cite{Chen:2023wpj} were $\Xi_0 =1.67^{+0.93}_{-0.94}$ and $n=0.8^{+3.59}_{-0.69}$. The above parametrization is based on empirical fits of the theoretical predictions for different modified gravity theories.

Another parametrization used in analyzing GWs data is 
\be
    r_d(z)=\exp \left\{\frac{c_M}{2\Omega_{\Lambda,0}} \ln \frac{1+z}{[\Omega_{m,0}(1+z)^3+\Omega_{\Lambda,0}]^{1/3}} \right\}\,, \label{Cm}
\ee
for which the best fit  value obtained in \cite{Chen:2023wpj} was $c_M=1.5^{+2.2}_{-2.1}$.
The above parametrization is based on assuming luminal gravitational wave propagation and a varying effective Planck, according to the following propagation equation
\begin{equation}
    \tilde{h}'' + {\cal H}[2+\alpha_M(z)]\tilde{h}'+c^2k^2\tilde{h} = 0,
\end{equation}
where $\alpha_M$ is defined by
\begin{equation}
    \alpha_M(z) \equiv \frac{d\ln(M_{\rm eff}/M_{\rm P})^2}{d\ln a}.
\end{equation}
In the EFT notation introduced above we have $M_{\rm eff}=\MM \sqrt{f}$.
Eq.(\ref{Cm}) is obtained by assuming this form for $\alpha_M$
\begin{equation}
    \alpha_M(z) = c_M \frac{\Omega_\Lambda(z)}{\Omega_{\Lambda,0}} = c_M \frac{1}{\Omega_{m,0} (1+z)^3 + \Omega_{\Lambda,0}},
\label{eq:alphaM_cM}
\end{equation}
in which $c_M$ is a free parameter, and $\Omega_\Lambda(z)$ is the fractional dark energy density.
In both the analyses outlined above, the one based on $\alpha_M$ and the one using $\{\Xi_0,n\}$, an effective $\Lambda$CDM background was assumed.

\section{Effects of modified gravity on scalar perturbations}
The effects of modified gravity on scalar perturbations  give rise  to  a modification of the  Poisson's equations   \cite{Linder:2015rcz}
\bea 
& &   \nabla^{2} \Psi = 4\pi a^{2} G^{\Psi}_{\rm eff} \rho_m\,\delta_ m \,, \\  
& &   \nabla^{2} \Phi = 4\pi a^{2} G^{\Phi}_{\rm eff} \rho_m\,\delta_ m \,, \\
& &   \nabla^{2} (\Psi+\Phi) = 8\pi a^{2} G^{\Psi+\Phi}_{\rm eff} 
\rho_m\,\delta_m \,, 
\eea 
where the effective gravitational constant is \cite{Linder:2015rcz,Bloomfield:2012ff}
\be 
\frac{G^{\Phi}_{\rm eff}}{G_N}= \frac{2M_p^2}{M_\star^2} 
\frac{[\alpha_B(1+\alpha_T)+2(\alpha_M-\alpha_T)]+\alpha_B'}{(2-\alpha_B)[\alpha_B(1+\alpha_T)+2(\alpha_M-\alpha_T)]+2\alpha_B'} \,, \label{eq:geff}
\ee 
and the gravitational slip $\bar\eta$ is 
\be
\bar\eta=
\frac{(2+2\alpha_M)[\alpha_B(1+\alpha_T)+2(\alpha_M-\alpha_T)]+(2+2\alpha_T)\alpha_B'}{(2+\alpha_M)[\alpha_B(1+\alpha_T)+2(\alpha_M-\alpha_T)]+(2+\alpha_T)\alpha_B'} \,.\label{eq:etafull} 
\ee 
The above equations have been derived by applying the EFT to Horndeski theories, we denote with prime  $d/d\,\rm{ln} \,a$, with a the scale factor, and we use the definition of gravitational slip $\bar\eta$ \cite{Bellini:2014fua}
\be 
\bar\eta=\frac{2\Psi}{\Psi+\Phi}=\frac{\geffs}{\geffsh} \label{bareta}
\,,
\ee
which is related to the other definition of slip by 
\bea
\eta&=&\frac{\Psi}{\Phi}=\frac{G^{\Psi}_{\rm eff}}{G^{\Phi}_{\rm eff}}=\frac{\bar\eta}{2-\bar\eta}\,.
\eea
The slip parameter can be estimated from lensing observations, since $G^{\Phi}_{\rm eff}$ governs the deflection of light, and for this reason is sometime denoted  as \cite{Linder:2016wqw} as $G_{\rm light}$.

From the above equations we get
\bea
G^{\Psi}_{\rm eff}&=&\frac{\bar\eta}{2-\bar\eta}G^{\Phi}_{\rm eff}=\eta\, G^{\Phi}_{\rm eff} \,,\\
\geffsh&=&\bar\eta \,G^{\Psi}_{\rm eff} =\frac{G^{\Psi}_{\rm eff}+G^{\Phi}_{\rm eff}}{2}=\frac{1}{2-\bar\eta}G^{\Phi}_{\rm eff}=\frac{1+\eta}{2}G^{\Phi}_{\rm eff} \,,
\eea
which are useful to compare to observations, since the quantity $G^{\Psi}_{\rm eff}$ is the one related to the growth of structure, while ${\geffsh}$ is related to the deflection of light \cite{Linder:2020xza}.
The relation between the coefficients of the EFT action and the property functions $\alpha_i$ can be found in \cite{Linder:2015rcz}. The speed of GWs is given by
\be
1+\alpha_T=v^2_{\rm GW}\,,\label{cTalpha}
\ee 
and the quantity $M_*$ is defined as \cite{Linder:2015rcz}
\be
M_*^2=\frac{M^2_{P}f}{1+\alpha_T}=\frac{\Omega^2}{v^2_{\rm GW}}\,,\label{ML}
\ee
so that the general relativity limit corresponds to $\{f=1,\alpha_i=0\}$, implying $G^{\Phi}_{\rm eff}/G_N=G^{\Psi}_{\rm eff}/G_N=\bar\eta=\eta=1$.

\subsection{Observational constraints}

Following the notation adopted in \cite{Ishak:2024jhs}, the EFT action can be  re-written as
\begin{equation}
  \begin{aligned}
S_{\text{DE}} &= \int d^4 x \sqrt{-g} \bigg[ M_{\text{P}}^2 [1+\Omega(t)] \frac{R}{2} - \Lambda(t) - c(t) g^{00} \\[0.2cm]
&+\frac{M_{2}^{4}(t)}{2}(\delta g^{00})^2 - \bar{M_{1}}^3(t) \frac{1}{2} \delta g^{00} \delta K - \bar{M_{2}}^2(t) \frac{1}{2} (\delta K)^2  \\[0.2cm]
&- \bar{M_3}^2(t) \frac{1}{2} \delta K{_\nu^\mu} \delta K{_\mu^\nu}  + \hat{M}^2(t) \frac{1}{2} \delta g^{00} \delta R^{(3)}+m_2(t) \partial_i g^{00}\partial^i g^{00} \bigg]\,, 
  \label{eq:EFTDEaction}
  \end{aligned}
\end{equation}
where we denote with $\delta R^{(3)}$  the perturbation of the spatial part of the Ricci scalar. 
It is convenient to introduce the dimensionless functions
\begin{equation}
    \begin{aligned}
          \gamma_1 &= \frac{M_{2}(t)^4}{m^2_0 H^2_0}, &
    \gamma_2 &= \frac{\bar{M}_{1}(t)^{3}}{m^2_0 H_0}, &
    \gamma_3 &= \frac{\bar{M}_{2}(t)^{2}}{m^2_0}, \\
    \gamma_4 &= \frac{\bar{M}_{3}(t)^2}{m^2_0}, &
    \gamma_5 &= \frac{\hat{M}(t)^{2}}{m^2_0}, &
    \gamma_6 &= \frac{m_{2}(t)^{2}}{m^2_0},
    \end{aligned}
    \label{eq:gamma}
\end{equation}
in terms of which the conditions  to avoid higher-order spatial derivatives take the form  
\be
2\gamma_5 = \gamma_3 = -\gamma_4 \quad,\quad \gamma_6 = 0. \label{gamma0} 
\ee
In \cite{Ishak:2024jhs} the following parametrization was adopted
\begin{equation}\label{eq:param_EFT_basis}
    \Omega(a) = \Omega_0 a^{s_0}\,, 
\end{equation}
where the parameter $s_0$ controls how early $\Omega(a)$ tend to GR prediction, and $\Omega_0$ is related to the strength of the gravitational coupling.  The  parameter $\Omega(a)$ is related to  $\alpha_M$ by
\begin{equation}
    \alpha_M = \frac{a}{\Omega + 1}\frac{d\Omega}{da}.
\end{equation} The $\gamma_1$ is related to kineticity in the EFT Lagrangian and $\gamma_2$  to the kinetic braiding. Both were set to zero in the analysis performed in \cite{Ishak:2024jhs}. 

The GWs speed is related to $\gamma_3$ by
\begin{equation}
v_{\rm GW}^2 = 1 - \frac{\gamma_3(a)}{1 + \Omega(a) + \gamma_3(a)}\,,
\end{equation}
and motivated by the GW170817 multimessenger event in \cite{Ishak:2024jhs} it was set $\gamma_{3}(a) =0$, which because of Eq.(\ref{gamma0}) implies  also $\gamma_4=\gamma_5=\gamma_6=0$.

The joint analysis  performed in \cite{Ishak:2024jhs}, using  data from the dark energy survey instrument (DESI), including full shape (FS) spectrum and baryon acoustic oscillations(BAO), cosmic microwave background (CMB) with lensing, and the five-year SN~Ia (DESY5SN) sample from the dark energy survey (DES), gives the constraints $\{\Omega_0 = 0.01189^{+0.00099}_{-0.012},s_0 = 0.996^{+0.54}_{-0.20}\}$.


Furthermore, the combined analysis of data from DESI (FS + BAO) with CMB with no-lensing, weak lensing and galaxy clustering  from DESY3 ($3\times 2$-pt), gives the constraints $\{ \Omega_0 =  0.0150^{+0.0041}_{-0.016},
s_0 = 1.06^{+0.49}_{-0.15}\}$.
Both set of constraints are consistent with GR, which corresponds to $\Omega(a) = 0$, and more specifically to $\Omega_0=0$. 

A similar analysis, but using the $\alpha$-basis, was performed in \cite{Ishak:2024jhs} assuming the parametrization
\begin{equation}\label{eq:alphas}
    \alpha_i(a) = c_i~\Omega_\mathrm{DE}(a)~,
\end{equation}
where $i = \{M, B, K, T\}$, $\Omega_\mathrm{DE}(a) \equiv \frac{8\pi G}{3H^2(a)}\rho_{\rm DE}(a)$, and the $c_i$ are  constant free parameters corresponding to the different property functions. The results of the joint analysis of data from DESI (FS+BAO), DESY5SN and CMB give the constraints 
\be
\{c_M = 1.05\pm 0.96, c_B = 0.92\pm 0.33\}.\label{ci}
\ee

\section{Consistency relations}
The observable quantities $\{r_d,G_{\rm eff},\bar\eta,v_{\rm GW}\}$ depend on the three functions \{$\alpha_T,\alpha_B,f\}$, or equivalently $\{\alpha_T,\alpha_B,\alpha_M\}$, since $\alpha_M$ is related to the derivative of $M_*$, so it is not really independent. 
In general it is not possible to obtain a consistency relation relating directly the observable quantities, without solving a differential equation, due to the presence of the derivative term $\alpha_B'$, but in some limits it is possible.  The function $\alpha_B$ is more difficult to obtain in terms of observables quantities, unless some extra conditions are imposed \cite{Linder:2020xza}.
In general the functions $f$ and $\alpha_T$ can be  obtained from $r_d$ and $v_{\rm GW}$, and then $\alpha_M$ is derived by taking the derivative of eq.(\ref{ML}).

In GR the GW and EMW luminosity distance is predicted to be the same, while in modified gravity theories they can differ, as shown in Eq.(\ref{rd}). For bright sirens the EMW luminosity distance can be estimated directly from the electromagnetic counterpart redshift, to obtain the distance ratio. For bright sirens it is also possible to measure the time delay between the detection of GWs and the associated EMWs, allowing to set constraints on the difference between the speed of GWs and EMWs.
For dark sirens the GW-EMW distance ratio can be obtained using statistical methods to infer the GW event redshift, using a combination of the  spectral siren and catalog methods. The spectral siren methods is assuming that the source frame masses follow the same distribution at any redshift, and that the difference between the source and detector frame masses is due to the GW frequency redshift. In the catalog method the probability of GW events is assumed to be proportional to the luminosity of the host galaxy, and a line of sight redshift prior is obtained from galaxy catalogs and used to infer the redshift of the events.  

\subsection{No-slip luminal (NSL) theories : $\bar\eta=1, \alpha_T=0$}
This is the case studied in \cite{Linder:2015rcz}. We report it because it is useful to check the consistency with the limits of the other cases.
From imposing the conditions $\{\bar\eta=1, \alpha_T=0\}$ in Eq.(\ref{eq:etafull}) we get $\alpha_B=-2 \alpha_M$, which substituted in Eq.(\ref{eq:geff}) gives
\be
8 \pi M_p^2\,G^{\Phi}_{\rm eff}=8 \pi M_p^2\,G^{\Psi}_{\rm eff}=\left[\frac{d^{\rm GW}_{\Lr}(z)}{  d^{\rm EM}_{\Lr}(z)}\right]^2 \label{GeffNS}\,,
\ee
where the first equality is a consequence of the fact that $\bar\eta=1$ implies $\eta=1$.

\subsection{No-braiding (NB) theories : $\alpha_B= \alpha'_B=0$}

In the no braiding (NB) limit, in which $\alpha_B= \alpha'_B=0$ , we get
\be
8 \pi M_p^2\, G^{\Phi}_{\rm eff} = 
 \frac{\alpha_T+1}{f }\,,\label{GeffNBa}
\ee
which using eq.(\ref{rd}) can be expressed in terms of observable quantities as
\be
8 \pi M_p^2\,G^{\Phi}_{\rm eff} =  8 \pi M_p^2\,\frac{2-\bar\eta}{\bar\eta}\,G^{\Psi}_{\rm eff} =\left[\frac{d^{\rm GW}_{\Lr}(z)}{  d^{\rm EM}_{\Lr}(z)}\right]^2\frac{{v_{\rm GW}(z)}}{{v_{\rm GW}(0)}}\,.\label{GeffNB}
\ee
Note that the condition $\alpha_B=0$ does not imply no-slip, since we have
\be
\bar\eta=\frac{2 \alpha_M+2}{\alpha_M+2}\,.
\ee
In the no-slip luminal limit Eq.(\ref{GeffNB}) reduces to Eq.(\ref{GeffNS}).
This consistency relation is in agreement with and generalize the results obtained in some luminal theories  of modified gravity such as no-slip luminal Horndeski theories \cite{Linder:2018jil} and some non local theories \cite{Belgacem:2017ihm}.
The no-brading luminal theories are called 'only run' in \cite{Linder:2020xza}. Note that the CR in Eq.(\ref{GeffNB}) is also satisfied by general relativity (GR), since in this case $v_{\rm GW}=\bar\eta=1$, $G_N=1/8 \pi M_p^2$, and $d^{\rm GW}_{\Lr}=d^{\rm GW}_{\Lr}$. This is expected, since GR is a no-brading theory. The left hand side (l.h.s) of the CR is related to larges scale structure observations, while the right hand side (r.h.s.) to gravitational waves observations. The no brading condition has some important impact on the stability of the theory, since  it is related to the early universe value of the sound speed \cite{Linder:2020xza}, so no brading theories may require extra conditions on $\alpha_M$ to ensure stability.

\subsection{Constant  brading (CB)
theories : $\alpha'_B=0$}
Setting $\alpha_B'=0$ in  Eq.(\ref{eq:etafull}) and Eq.(\ref{eq:geff}) we get
\bea
\bar\eta&=&\frac{2 (\alpha_M+1)}{\alpha_M+2} \,,\label{etaCB} \\
8 \pi M_p^2\,G^{\Phi}_{\rm eff}&=&\frac{2 (\alpha_T+1)}{(2-\alpha_B) f }\,.
\eea
We can then express $f$ in terms of $r_d$ using Eq.(\ref{rd}), obtaining a relation in terms  of observable quantities only
\be
8 \pi M_p^2\, G^{\Phi}_{\rm eff}=8 \pi M_p^2\,  \frac{2-\bar\eta}{\bar\eta}G^{\Psi}_{\rm eff}=\frac{2}{2-\alpha_B(c_s,\bar\eta,v_{\rm GW})}\left[\frac{d^{\rm GW}_{\Lr}(z)}{  d^{\rm EM}_{\Lr}(z)}\right]^2 \frac{v_{\rm GW}(z)}{v_{\rm GW}(0)}\,. \label{GeffCB}
\ee
In the no-brading limit Eq.(\ref{GeffCB}) takes in Eq.(\ref{GeffNB}), and imposing the additional no-slip luminal condition it reduces to Eq.(\ref{GeffNS}). The parameter $\alpha_B$ can be obtained by in terms of the sound speed $c_{s}$ \cite{Bellini:2014fua}
\be
c_{s}^{2}=-\frac{\left(2-\alpha_B\right)\left[\mathfrak{H}-\frac{1}{2}H^{2}\alpha_B\left(1+\alpha_{\textrm{T}}\right)\right]-H\dot{\alpha}_{\textrm{B}}+\tilde{\rho}_{\textrm{m}}+\tilde{p}_{\textrm{m}}}{H^{2}D}\,,
\ee
by imposing $\alpha_B'=0$, and  solving an algebraic equation giving $\alpha_B(c_s,\bar\eta,v_{\rm GW})$.
In the above equation we have used the notation \cite{Bellini:2014fua}
\bea 
\mathfrak{H} &=&\dot{H}-H^{2}\left(\alpha_M-\alpha_{\textrm{T}}\right)\,,\label{eq:Upsilon}\\
D &=& \alpha_{\textrm{K}}+\frac{3}{2}\alpha_B^{2}\,.
\eea
The dependency of the CR on the sound speed and the GWs speed is expected, since the brading is related to a mixing between tensor and scalar perturbations.
Eq.(\ref{GeffCB}) establishes a relation between different observables which could be affected by gravity modification:  the electromagnetic and gravitational luminosity distances, the effective gravitational coupling, the slip and the speed of gravitational waves, and the sound speed.  Alternatively it can be considered a consistency relation between scalar and tensor perturbations. Note that constant brading is strongly  constrained by early universe observations such as the cosmic microwave background radiation \cite{Linder:2020xza}, so for practical applications the CR could be interesting only al low redshift.

\subsection{Power law brading (PLB) theories : $\alpha_B \propto \, a^b$}
A natural parametrization of $\alpha_B$ in cosmology is a power law of the scale factor 
\be
\alpha_B=\alpha_{B0}\, a^b \,, \label{abpower}
\ee
where $a$ is the scale factor. 
Substituting Eq.(\ref{abpower}) in Eq.(\ref{eq:etafull}) gives
\be
\bar\eta=\frac{2 \alpha_{B0} a^b(\alpha_T+1)   (\alpha_M+b+1)+4 (\alpha_M+1) (\alpha_M-\alpha_T)}{\alpha_{B0} a^b (\alpha_M \alpha_T+\alpha_M+\alpha_T b+2 \alpha_T+2 b+2)+2 (\alpha_M+2) (\alpha_M-\alpha_T)}\,.\label{etapl}
\ee
Solving the above equation for $\alpha_{B0}$ we obtain
\be
\alpha_{B0}=\frac{2 a^{-b} (\alpha_T-\alpha_M) [\alpha_M (\bar\eta -2)+2 (\bar\eta -1)]}{\alpha_M (\alpha_T+1) (\bar\eta -2)+\alpha_T (b\, \bar\eta -2 b+2 \bar\eta -2)+2 (b+1) (\bar\eta -1)}\label{b0}\,.
\ee
Note that in the constant brading limit corresponding to $b=0$, Eq.(\ref{etapl}) takes the expected form given in Eq.(\ref{etaCB}), which only depends on $\alpha_{M}$, implying that Eq.(\ref{b0}) is not valid in that limit.
 From Eq.(\ref{rd}) we get
\be
f(r_d,v_{\rm GW})= \frac{1}{r_d^2}\frac{v_{\rm GW}(a)}{v_{\rm GW}(a_0)} \,,\label{frd}
\ee
where $a_0$ denotes the value of the scale factor today.
We can then obtain $\alpha_M=(ln M_*^2)'$ in terms of observable quantities by combing Eq.(\ref{ML}) and Eq.(\ref{frd})
\be
\alpha_M(r_d,v_{\rm GW})=\frac{d}{d \,ln \,a}\left[ln\left(\frac{M_P^2 f}{v_{\rm GW}^2}\right)\right]\,.\label{amrv}
\ee
Substituting Eq.(\ref{abpower}) and Eq.(\ref{b0}) in Eq.(\ref{eq:geff}) we  obtain 
\be
8 \pi M_p^2\,G^{\Phi}_{\rm eff}=\frac{(\alpha_T+1) (\bar\eta -2) [(\alpha_M+2) (\alpha_T+1) \bar\eta -2 (\alpha_T+1) (\alpha_M+b+1) +(\alpha_T+2) b\, \bar\eta ]}{f  \{\bar\eta  [2 (\alpha_M+1)-(\alpha_M+2) \bar\eta ]+b (\bar\eta -2) [(\alpha_T+2) \bar\eta -2 (\alpha_T+1)]\}}\,, \label{GeffPL}
\ee
which combined with Eq.(\ref{frd}), Eq.(\ref{amrv}) and Eq.(\ref{cTalpha}), gives the effective gravitational constant  in terms of the observational quantities $r_d,f,\bar\eta$,$v_{\rm GW}$ and the brading exponent $b$.
Note that Eq.(\ref{GeffPL}), in NSL limit corresponding to \{$\alpha_T=0,\bar\eta=1\}$, reduces to Eq.(\ref{GeffNS}). Eq.(\ref{GeffPL}) is the most general CR derived in this paper, but in the constant brading limit is not valid, because as mentioned above, in that limit $\bar\eta$ is given by Eq.(\ref{etaCB}), and Eq.(\ref{b0}) is not valid. For this reason a separate derivation is necessary to obtain the constant brading CR given in Eq.(\ref{GeffCB}).
\section{General Horndeski multi-probe consistency condition}
In the previous sections we have focused on consistency conditions derived under different assumption for the function $\alpha_B$. This has allowed to derive compact analytical relations between observables such as the NB and CB consistency conditions given in eq.(\ref{GeffNB}) and eq.(\ref{GeffCB}).
If the function $\alpha_B$ is left completely free it is possible to derive a general consistency condition satisfied by all Horndeski theories, but this takes a less compact mathematical form.

The steps to estimate the effective gravitational coupling from GWs observations are the followings:
\begin{itemize}
    \item Obtain $f^{\rm GW}(r_d,v_{\rm GW})$ from the distance ratio using eq.(\ref{frd}) and from it $\alpha^{\rm GW}_M(r_d,v_{\rm GW})$ using eq.(\ref{amrv}). This can be done always, without making any assumption on the form of $\alpha_B$, because the distance ratio $r_d$ is independent of $\alpha_B$. 
\item Compute $G^{\rm GW}_{\rm eff}$ from GWs observations from eq.(\ref{eq:geff}) as
\be 
\frac{G^{\rm GW}_{\rm eff}}{G_N}= \frac{2 \,\eta \, v^2_{\rm GW}}{f^{\rm GW}(r_d,v_{\rm GW})} 
\frac{[\alpha_B\,v^2_{\rm GW}+2(\alpha_M^{\rm GW}(r_d,v_{\rm GW})v^2_{\rm GW}+1)]+\alpha_B'}{(2-\alpha_B)[\alpha_B \, v^2_{\rm GW}+2(\alpha_M^{\rm GW}(r_d,v_{\rm GW})-v^2_{\rm GW}+1)]+2\alpha_B'} \,. \label{GeffGw}
\ee
\end{itemize}
We are denoting quantities obtained from GWs observation with a $\rm GW$ superscript or subscript, and for ease of notation we have dropped the $\psi$ superscript, i.e. $G^{\rm GW}_{\rm eff}$ stands for $G^{\psi,\rm GW}_{\rm eff}$, and the quantities $\eta$ and $\alpha_B$ can be estimated from LSS data \cite{Ishak:2024jhs}. When assuming no-brading or constant brading the above equation reduces to the analytical consistency condition derived in the previous sections.
Another independent way to estimate the effective gravitational coupling is to analyze LSS data assuming the modified Poisson equation
\be
 \nabla^{2} \Psi = 4\pi a^{2} G^{\rm LSS}_{\rm eff} \rho_m\,\delta_ m \,,\label{PLSS} \\ 
\ee
and for any Horndeski theory we obtain the general consistency condition (GCR)
\be
\frac{G^{\rm GW}_{\rm eff}}{G^{\rm LSS}_{\rm eff}}=1  \,. \label{CRGEN}
\ee
The above equation is a consequence of the fact that  $G^{\rm GW}_{\rm eff}$ can be obtained from GWs observations using eq.(\ref{GeffGw}), and  this is theoretically expected to be the same as the quantity  $G^{\rm LSS}_{\rm eff}$ appearing in eq.(\ref{PLSS}), obtained from LSS data analysis.
The GCR should be satisfied by any Horndeski theory, since no assumption about the functional form of any function appearing in it has been made.

Eq.(\ref{CRGEN}) relates in a single equation five observables, two related to LSS and lensing observations, $\{\eta, G_{\rm eff}\}$, and  three related to GWs observations, $\{ d^{\rm GW}_{\Lr}(z), d^{\rm EM}_{\Lr}(z), v_{\rm GW}\}$, and it allows to test if Horndeski theories are consistent with these different observables under minimal assumptions about the parameterizations used to analyze data.

\section{Large scale structure observations implications for the GW-EMW luminosity distance}
Large scale structure observations can be used to constrain $G^{\Psi}_{\rm eff}$ and $G^{\Psi+\phi}_{\rm eff}$, and the recent DESI \cite{Ishak:2024jhs} results are setting  stringent constraints on their redshift dependence.
Assuming the GW speed to be the same as the speed of light, the consistency relation gives a relation between $G^{\Psi}_{\rm eff}$ and the GW-EMW distance ratio. This can be used to estimate what can be the expected deviation of GWs observations from GR for the theories satisfying the CRs.

\begin{figure}
\includegraphics{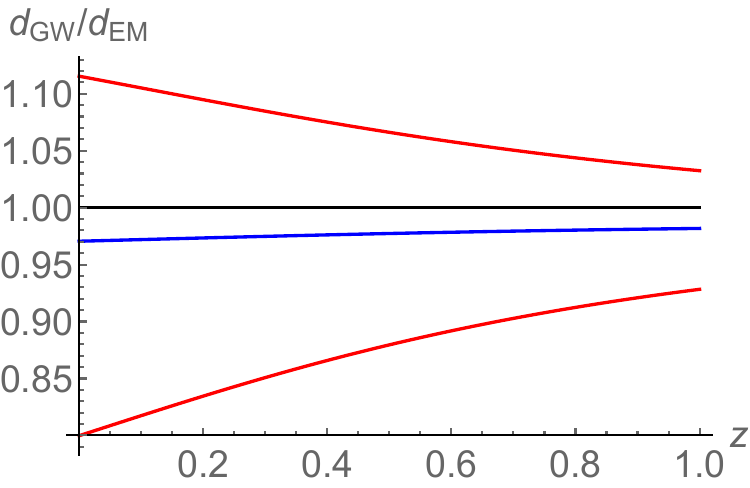}
\caption{The GW-EMW distance ratio implied by non GW observations is plotted in blue as a function of redshift, using the best fit parameters in Eq.(\ref{best}),obtained in \cite{Ishak:2024jhs}. The red lines are the $68\%$ confidence interval bands. This plot was obtained assuming luminal constant braiding theories with $\alpha_B=0.05$,  using the consistency condition in Eq.(\ref{GeffCB}). 
}
\label{dgwdemCB}
\end{figure}

\begin{figure}
\includegraphics{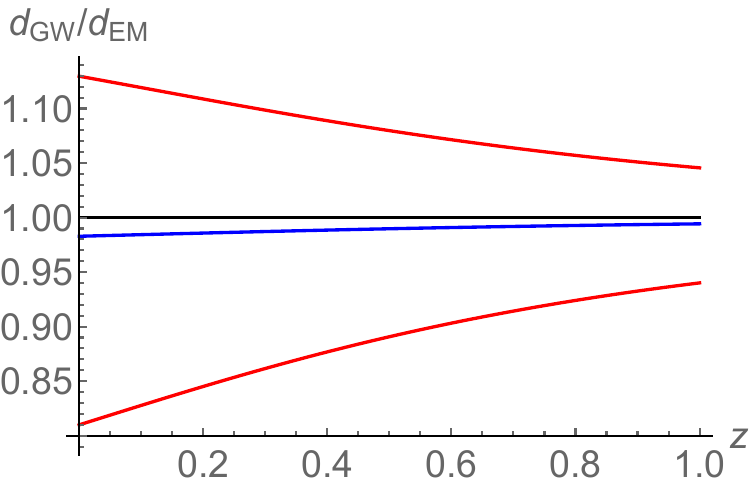}
\caption{The GW-EMW distance ratio implied by non GW observations is plotted in blue as a function of redshift, using the best fit parameters in Eq.(\ref{best}), obtained in \cite{Ishak:2024jhs}. The red lines are the $68\%$ confidence interval bands. This plot was obtained assuming luminal no-braiding theories, i.e. using Eq.(\ref{GeffNB}). 
}
\label{dgwdemNB}
 \end{figure}

For a general Horndeski theory obtaining the distance ratio in terms of LSS observations requires to solve a differential equation, because of the derivative operator in Eq.(\ref{amrv}), but for constant brading it is sufficient to solve an algebraic equation. From the constant brading consistency condition in Eq.(\ref{GeffCB}) we can in fact get an expression for the GW-EMW distance ratio in terms of LSS observations
\be
\frac{d^{\rm GW}_{\Lr}(z)}{  d^{\rm EM}_{\Lr}(z)} =\left[\frac{2-\bar\eta}{\bar\eta}\frac{2-\alpha_B}{2}\frac{v_{\rm GW}(0)}{v_{\rm GW}(z)} \frac{G^{\Psi}_{\rm eff}(z)}{G_ N}\right]^{1/2}\,. \label{rdGeffCB}
\ee
Using the parametrizations \cite{Ishak:2024jhs}
\bea
\frac{G^{\Psi}_{\rm eff}}{G_N}&=& \mu(a)= \left[1+\mu_0 \frac{\Omega_{\Lambda}(a)}{\Omega_{\Lambda}}\right]\nonumber\,,\\
\frac{G^{\Psi+\Phi}_{\rm eff}}{G_N}&=& \Sigma(a)= \left[1+\Sigma_0 \frac{\Omega_{\Lambda}(a)}{\Omega_{\Lambda}}\right]\label{musigma} \,,
\eea
the best fit values \footnote{  DESI+CMB(LoLLiPoP-HiLLiPoP)-nl+DESY3+DESSNY5 analysis} , assuming no scale dependence and a flat $\Lambda$CDM background, are 
\bea
\mu_0=0.05\pm 0.22 &,&\Sigma_0=0.008\pm 0.045\,.\label{best}
\eea
 From Eq.(\ref{bareta}) we get that for this parametrization $\bar\eta=\mu(a)/\Sigma(a)$.
Using the best fit parameters in Eq.(\ref{best}) we show in Fig.(\ref{dgwdemCB}) and Fig.(\ref{dgwdemNB})  the GW-EMW distance ratio implied by non GW observations for luminal constant brading and no-brading theories, corresponding respectively to Eq.(\ref{GeffCB}) and Eq.(\ref{GeffNB}). 

\begin{figure}
\includegraphics{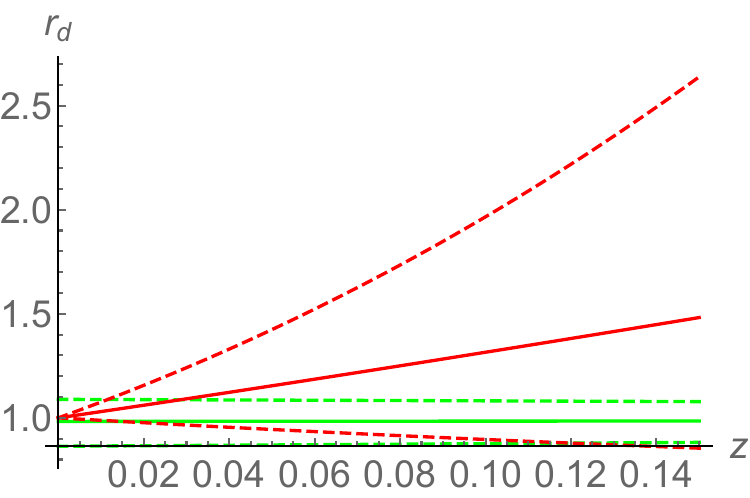}
\caption{The GW-EMW distance ratio implied by non GW observations is plotted in green as a function of redshift, using the best fit parameters in Eq.(\ref{best}), obtained in \cite{Ishak:2024jhs}. The distance ratio estimated from GWs observations assuming the parametrization given in eq.(\ref{Cm}) is plotted in red, using the best fit parameters estimated in \cite{Chen:2023wpj}. The dashed lines delimit the $68\%$ confidence interval bands. This plot was obtained assuming luminal no-braiding theories,  using the consistency condition in Eq.(\ref{GeffNB}). 
}
\label{rdcmGeffNB}
\end{figure}

\begin{figure}
\includegraphics{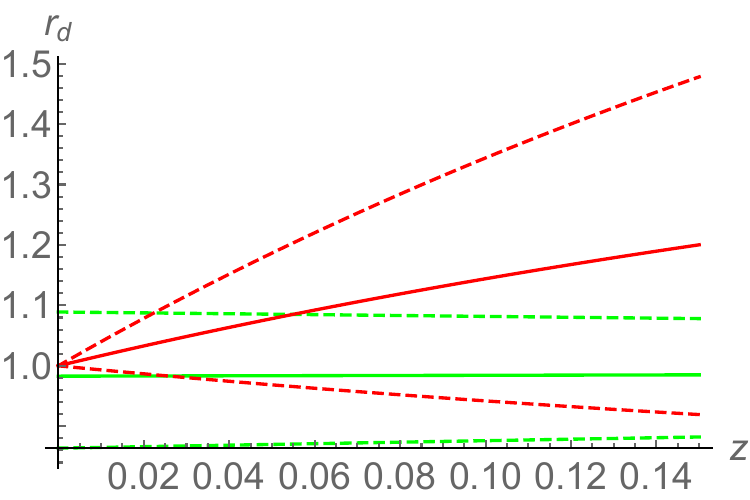}
\caption{The GW-EMW distance ratio implied by non GW observations is plotted in green as a function of redshift, using the best fit parameters in Eq.(\ref{best}), obtained in \cite{Ishak:2024jhs}. The distance ratio estimated from GWs observations assuming the parametrization given in eq.(\ref{Xi0}) is plotted in red, using the best fit parameters estimated in \cite{Chen:2023wpj}. The dashed lines delimit the $68\%$ confidence interval bands. This plot was obtained assuming luminal no-braiding theories,  using the consistency condition in Eq.(\ref{GeffNB}). 
}
\label{rdX0GeffNB}
\end{figure}

In Fig.(\ref{rdcmGeffNB}) and Fig.(\ref{rdX0GeffNB}) we show a comparison between the distance ratio constraints from GWs observations \cite{Chen:2023wpj} and those implied by LSS observations using Eq.(\ref{rdGeffCB}), assuming luminal no-brading theories.
The constraints are compatible at the present level of experimental uncertainty, confirming the validity of the no-brading assumption, and showing that LSS observations imply tighter constraints.

\section{Generalized phenomenological consistency condition}
Inspired by Eq.(\ref{GeffNB}) we propose a generalized phenomenological consistency condition of the form
\be
 8 \pi M_p^2\,\left(\frac{2-\bar\eta}{\bar\eta}\right)^{n_{\bar\eta}}\,G^{\Psi}_{\rm eff} =8 \pi M_p^2\,\eta^{n_{\eta}}\,G^{\Psi}_{\rm eff} =(1+\alpha_B)^{n_B}\left[\frac{d^{\rm GW}_{\Lr}(z)}{  d^{\rm EM}_{\Lr}(z)}\right]^{n_d}\left[\frac{{v_{\rm GW}(z)}}{{v_{\rm GW}(0)}}\right]^{n_v} \label{GeffGen}\,,
\ee

where the phenomenological parameters $\{n_{\bar\eta}=-n_{\eta},n_d,n_v\}$ in  general relativity (GR) take the values corresponding to Eq.(\ref{GeffNB}) $\{n_{\bar\eta}=1,n_d=2,n_v=1\}$, since GR is  a no-brading theory, and Eq.(\ref{GeffNB}) is valid for any no-brading theory. For constant small brading theories we have $\{n_{\bar\eta}=1,n_d=2,n_v=1,n_B\approx1/2\}$. Note that Eq.(\ref{GeffGen}) in GR is satisfied by any set of $n_i$, since in this case all the arguments of the power laws have unitary value.
In terms of the $\{\Sigma,\mu\}$ parametrizations given in Eq.(\ref{musigma}) the generalized CR takes the form
\bea
\left(\frac{2\Sigma-\mu}{\mu}\right)^{n_{\bar\eta}}\mu=(1+\alpha_B)^{n_B} \left[\frac{d^{\rm GW}_{\Lr}(z)}{  d^{\rm EM}_{\Lr}(z)}\right]^{n_d}\left[\frac{{v_{\rm GW}(z)}}{{v_{\rm GW}(0)}}\right]^{n_v}\,,
\eea
which is satisfied in GR, since in this case $(2\Sigma-\mu)/\mu=\mu=1$.
In general the parameters $n_i$ could have a time dependence, i.e. we could have $n_i(z)$.

\section{Conclusions}
We have derived  a set of consistency conditions for different classes of Horndeski theories defined by the behavior of the function $\alpha_B$, relating directly the effective gravitational constant, the slip parameter, the GW and EMW luminosity distance, the GW speed and the sound speed. The no-braiding consistency relation is also satisfied by GR, and can be used as a multi-probe test of general relativity. These consistency conditions relate observable quantities directly, and do not depend on the parametrization adopted to analyze observational data. They are hence particularly suited for model independent data analysis, or to check the consistency between the results obtained assuming different parameterizations. 
A general consistency condition has also been derived, satisfied by any Horndeski theory for any form of $\alpha_B$, allowing to test the compatibility of Horndeski theories with LSS and GWs.

Some examples of the applications of the consistency conditions to compare  observational constraints obtained from GWs and LSS have been given, showing for example that under the assumption of luminality and no-brading, the effective gravitational coupling inferred form GWs observations is consistent with the constraints obtained from LSS observations.
In the future it will be interesting to perform a joint analysis of large scale structure data and GW observations to verify the validity of the different CRs.
Inspired by the form of the CRs for no-brading and constant brading theories, we have also proposed a generalized phenomenological consistency condition, which could be used for model independent observational data analysis, without assuming any specific class of theory.

A violation of the CRs would imply  a violation of the assumptions made for the property function $\alpha_B$. In the case of the general consistency condition a violation would imply that observational data are not consistent with Horndeski  theories, within the limits of the approximations used to derive Eq.(\ref{eq:geff}).
While in this paper we have considered the effects of modified gravity on the lensing of electromagnetic signals, in the future it will be interesting to extend this analysis to  gravitational waves lensing \cite{Ezquiaga:2020dao,Ezquiaga:2022nak}.

Since the GW strain is inversely proportional to the GW luminosity distance, while the apparent magnitude of galaxies is inversely proportional to the square of the electromagnetic luminosity distance, the CRs could be used in the future to obtain high redshift estimations of the effective gravitational constant and slip parameter using GW events with an EM counterpart at distances where large scale structure observations are not available or are not very precise, due to selection effects.

\section{Acknowledgments}
I thank Hsu Wen Chiang, Eric Linder, Sergio Vallejo, Johannes Noller and Tessa Baker for useful comments and discussions, and the Academia Sinica and HCWB for the kind hospitality.

\appendix

\section{Constant brading running theories : $\alpha_B'=const$}
In this section we derive a consistency condition for constant brading running theories satisfying the condition $\alpha_B'=const$. 
Setting $\alpha_B'=b$, where $b$ is a constant, we get
\be
\alpha_B=b\, ln(a)\,,
\ee
which substituted in Eq.(\ref{eq:etafull}) gives
\be
\bar\eta=\frac{2 (\alpha_M+1) [(\alpha_T+1) b \log (a)+2 (\alpha_M-\alpha_T)]+2 (\alpha_T+1) b}{(\alpha_M+2) [(\alpha_T+1) b \log (a)+2 (\alpha_M-\alpha_T)]+(\alpha_T+2) b}\,.
\ee
Solving the above equation for $b$ we obtain
\be
b=\frac{2 (\alpha_T-\alpha_M) [\alpha_M (\bar\eta -2)+2 (\bar\eta -1)]}{(\alpha_T+1) \log (a) [\alpha_M (\bar\eta -2)+2 (\bar\eta -1)]+\alpha_T (\bar\eta -2)+2 (\bar\eta -1)}\,. \label{b}
\ee
Substituting $\alpha_B$ in Eq.(\ref{eq:geff}) we  obtain
\be
8 \pi M_p^2\,G^{\Phi}_{\rm eff}=\frac{v_{\rm GW}^2 (\bar\eta -2) \{v_{\rm GW}^2 \log (a) [(\alpha_M+2) \bar\eta -2 (\alpha_M+1)]+(v_{\rm GW}^2+1) \bar\eta -2 v_{\rm GW}^2\}}{\bar\eta\,  f  \log (a) [2 (\alpha_M+1)-(\alpha_M+2) \bar\eta ]+f  \,(\bar\eta -2)  [(v_{\rm GW}^2+1) \bar\eta -2 v_{\rm GW}^2]}\,,
\ee
which combined with Eq.(\ref{frd}) and Eq.(\ref{amrv}) gives the effective gravitational constant only in terms of the observational quantities $r_d,f,\bar\eta$ and $v_{\rm GW}$.

\section{Friedman equations}
The modified Friedman equations are \cite{Gubitosi:2012hu}
\begin{align}
 H^2 + \frac{k}{a^2}    &=  \frac1{3 f \MM^2} (\rho_m + \rho_{D}  )  \label{frie1}\; ,\\
\dot H -  \frac{k}{a^2} &=  - \frac1{2 f \MM^2} (\rho_m + \rho_{D} +p_m + p_{D}  )  \label{frie2}\;.
\end{align}
After defining $p_D^{\rm eff}$ and $\rho_D^{\rm eff}$ according to
\begin{equation}
\rho_D = f\rho_D^{\rm eff}+ (f-1) \rho_m  , \qquad p_D = f p_D^{\rm eff} + (f-1) p_m \, .
\end{equation}
 eqs.~(\ref{frie1}-\ref{frie2}) take  a form similar to the one in general relativity
\be
 H^2 + \frac{k}{a^2}    =  \frac1{3 \MM^2} (\rho_m + \rho^{\rm eff}_{D}  )  \; ,\qquad
\dot H -  \frac{k}{a^2} =  - \frac1{2  \MM^2} (\rho_m + \rho^{\rm eff}_{D} +p_m + p^{\rm eff}_{D}  )  \;.
\ee
The advantage of the second form is that it allows to fix the background to a fiducial $\Lambda$CDM model, which allows a minimal change in the existing numerical codes designed assuming general relativity.


\bibliographystyle{h-physrev4}
\bibliography{mybib}
\end{document}